\documentclass[twocolumn,prl]{revtex4}

\usepackage{epsfig}
\usepackage{dcolumn}
\usepackage{bm}

\begin{document}

\preprint{APS/123-QED}

\title{Disorder   induced critical phenomena   in
magnetically glassy Cu-Al-Mn alloys}

\author{Jordi Marcos$^1$}
\author{Eduard Vives$^1$}
\email{eduard@ecm.ub.es}
\author{Llu\'{\i}s Ma\~{n}osa$^1$}
\author{Mehmet Acet$^2$}
\author{Ey\"up Duman$^2$} 
\author{Michel Morin$^3$} 
\author{V\'aclav Nov\'ak$^4$}
\author{Antoni Planes$^1$}

\affiliation{
$^1$ Departament  d'Estructura   i   Constituents  de   la
Mat\`eria,  Universitat  de Barcelona, Diagonal  647, Facultat  de
F\'{\i}sica, 08028 Barcelona, Catalonia\\
$^2$ Tieftemperaturhysik, Gerhard-Mercator Universit\"at, 47048 Duisburg, 
Germany\\
$^3$ Groupe d'Etudes de M\'etallurgie Physique et Physique des Mat\'eriaux,
INSA de Lyon, 20, Av.\ A.\ Einstein, 69621 Villeurbanne, France
$^4$ Institute of Physics, Na Slovance 2, Prague 8, 18221 Czech Republic
}

\date{\today}

\begin{abstract}
Measurements  of  magnetic  hysteresis  loops in  Cu-Al-Mn  alloys  of
different Mn content at low temperatures are presented.  The loops are
smooth  and continuous  above  a certain  temperature,  but exhibit  a
magnetization discontinuity below  that temperature.  Scaling analysis
suggest that this system  displays a disorder induced phase transition
line.   Measurements   allow  to  determine   the  critical  exponents
$\beta=0.03\pm 0.01$  and $\beta  \delta = 0.4  \pm 0.1$  in agreement
with those  reported recently [Berger  et al., Phys. Rev.   Lett. {\bf
85}, 4176 (2000)].
\end{abstract}

\pacs{Valid PACS appear here}

\maketitle

Beyond equilibrium, the  response of a physical system  to an external
applied force  shows hysteresis. Close to the  rate independent limit,
when thermal fluctuations  do not control the kinetics  of the system,
the hysteresis path is reproducible from cycle to cycle. This athermal
behavior  is observed,  for  instance, in  the magnetization  reversal
through  a first-order  transition  in many  magnetic  systems at  low
enough temperatures.  Modeling the shape  of the hysteresis loops is a
difficult problem that has attracted a great deal of interest for many
years. Theoretical studies of the zero-temperature (absence of thermal
fluctuations) hysteresis in a  number of lattice models with different
kinds  of quenched disorder  (random fields  \cite{Sethna1993}, random
bonds\cite{Vives1994},  random  anisotropies\cite{Vives2001},  etc...)
have shown that by increasing the strength of disorder, the hysteresis
loops change from displaying a discontinuous jump in the magnetization
reversal (ferromagnetic character) to behave smoothly with the applied
field.   This  change  has  been  interpreted as  a  disorder  induced
non-equilibrium phase transition.  For this phase transition, critical
exponents corresponding  to different  models have been  computed from
numerical  simulations and  renormalization group  methods.  Moreover,
universal scaling  has been predicted.  For  instance, for short-range
$3d$    models\cite{Perkovic1999}     $\beta=    0.035\pm0.028$    and
$\beta\delta=     1.81\pm0.32$,     whereas     for     mean     field
models\cite{Dahmen1996} $\beta= 1/2$ and $\beta\delta= 3/2$.

In contrast  to the considerable amount of  theoretical research, only
very  few experimental  works  have been  devoted  to corroborate  the
actual  existence  of  disorder   induced  phase  transitions  in  the
hysteresis loop.   Obrad\'o et al.  \cite{Obrado1999}  showed that the
hysteresis loops  in the magnetically glassy phase  of Cu-Al-Mn change
with  increasing  disorder  (related  to  the  alloy  composition)  as
predicted  by  the  models.    However,  no  critical  exponents  were
estimated  in  that work.   Later,  Berger  et al.   \cite{Berger2000}
studied the  effect of  magnetic disorder on  the reversal  process in
Co/CoO bilayers.  In this case, the antiferromagnetic CoO layer allows
a reversible tuning of  disorder by temperature variation.  They prove
the  existence of hysteresis-loop  criticality and  determine critical
exponents.   Nevertheless, the obtained  exponents ($\beta  = 0.022\pm
0.006$   and  $\beta  \delta   =  0.30   \pm  0.03$)   show  important
discrepancies  with  the  theoretical  predictions.   Therefore,  more
effort is needed  in the experimental analysis of  other candidates in
order to  confirm the existence of disorder  induced phase transitions
as predicted by theory.

In  this letter, we  present magnetic  hysteresis loops  for different
compositions  and temperatures  below  the freezing  temperature in  a
Cu-Al-Mn alloy.  A critical  transition line associated with the onset
of    a    discontinuity   in    the    hysteresis    loop   in    the
temperature-field-composition  space  has  been  determined,  and  the
critical  exponents  have  been  obtained  from  the  scaling  of  the
magnetization curves.  Our exponents  show a remarkable agreement with
the previous experimental estimates \cite{Berger2000}, thus supporting
the  idea  of  universality  in  such  disorder  induced  transitions.
Nevertheless, results also indicate  that $T=0$ theoretical models may
lack an essential  physical ingredient in order to  account for finite
temperature measurements.

A Cu$_{66.6}$Al$_{20.4}$Mn$_{13.0}$  single crystal has  been grown by
the  Bridgeman  method.   The  crystallographic  structure  is  $L2_1$
($Fm3m$)which can  be viewed  as four interpenetrated  fcc sublattices
\cite{Obrado1999}.   From  the ingot,  a  parallelepiped specimen  ($a
\simeq 3.4$ mm, $b \simeq 2.4$ mm  and $c \simeq 2.2$ mm) has been cut
with  $a$ and  $b$  oriented  along the  [001]  and [110]  directions,
respectively.  The sample was annealed  for 10 min at 1080 K, quenched
in a mixture of ice and water and aged at room temperature for several
days.   The  magnetic  properties  of Cu-Al-Mn  arise  from  localized
magnetic moments  at Mn-atoms.  These  moments are coupled  through an
oscillating  effective RKKY  interaction.   Due to  non-stoichiometry,
Mn-atoms are not uniformly distributed on one of the $L2_1$ sublattice
sites.  This disorder is at the origin of the glassy behavior observed
at low temperatures \cite{Obrado1999b}.
\begin{center}
\begin{figure}[thb]
\includegraphics[width=7cm,clip]{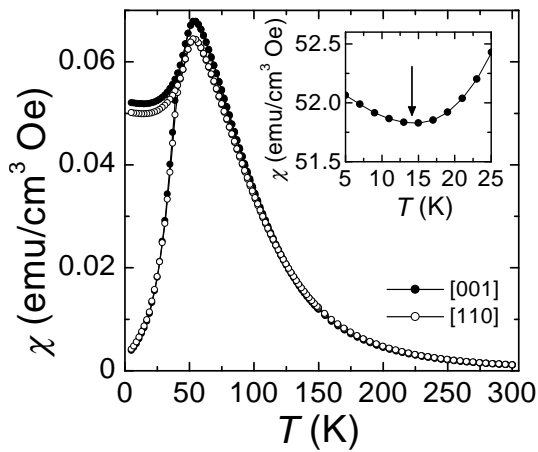}
\caption{\label{fig1} ZFC and FC low-field susceptibility with $H=100$
Oe along  the [001]  ($\bullet$) and [110]  ($\circ$) crystallographic
directions.   The inset  reveals the  existence of  a minimum  at $T_c
\simeq 15$ K in the FC curve.}
\end{figure}
\end{center}
\vspace{-9mm}
The magnetization as a function  of temperature and magnetic field has
been measured with a  SQUID magnetometer.  Results presented here have
been corrected for demagnetizing field effects. Fig.\ \ref{fig1} shows
the  behavior of the  zero field  cooled (ZFC)  and field  cooled (FC)
susceptibilities as a function of temperature.  Measurements have been
done by applying a field  of 100 Oe.  The different symbols correspond
to  the  field  applied   in  the  [001]  and  [110]  crystallographic
directions.  No  significant anisotropy is  observed.  The ZFC  and FC
curves split  at a  temperature of  $43$ K below  the position  of the
$\chi$  peak at  $\sim 55$  K.  The  splitting confirms  the  onset of
irreversibility   and  the  glassy   behavior  at   low  temperatures.
Moreover, the inset shows an enlarged view of the FC susceptibility in
this  glassy  region.   The  existence  of a  minimum  in  this  curve
indicates that,  along this metastable  path, there is a  tendency for
the system to  develop a field induced ferromagnetic  component at low
temperatures.  No indication of such a tendency can be observed in the
ZFC curve.

\begin{center}
\begin{figure}[thb]
\includegraphics[width=7cm,clip]{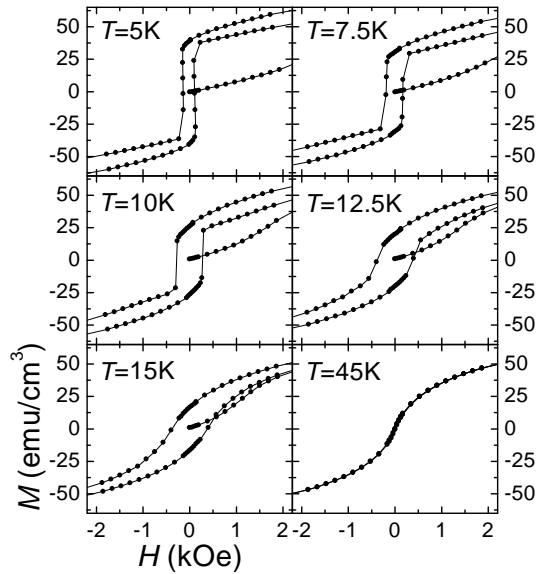}
\caption{\label{fig2}  Central  part   of  the  hysteresis  cycles  at
different temperatures  (as indicated)  obtained after cooling  in the
absence  of an external  field and  slowly varying  the field  from $0
\rightarrow 50$ kOe$ \rightarrow -50$ kOe $ \rightarrow 50$ kOe.}
\end{figure}
\end{center}

\begin{center}
\begin{figure}[thb]
\includegraphics[width=7.3cm,clip]{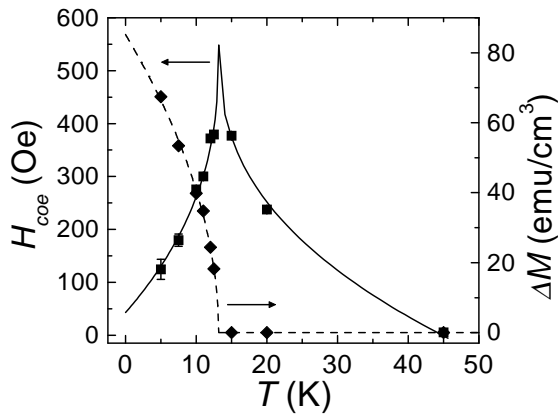}
\caption{\label{fig3} Behavior of the coercive field $H_{coe}$ and the
magnetization  jump  $\Delta M$  as  a  function  of temperature.  The
continuous line is  a fit of the integral  of equation \ref{clausius},
and the discontinuous line is a guide to the eyes.}
\end{figure}
\end{center}

Fig.\ \ref{fig2}  shows a series  of examples of hysteresis  cycles at
selected temperatures  in the range from  $5$K up to  $45$K. Each loop
has  been obtained  by  cooling from  $150$  K in  the  absence of  an
external field down to the selected temperature and slowly varying the
field from  $0 \rightarrow 50$ kOe$ \rightarrow  -50$ kOe$ \rightarrow
50$ kOe.  This allows to obtain the virgin magnetization curve and the
full saturation  loop.  Although  it is not  displayed in  the figure,
above  $40$  kOe  (and below  $-40$  kOe)  the  loops show  already  a
reversible behavior.  Nevertheless, saturation of the magnetization is
still not  reached at  $50$ kOe.   To a high  degree of  accuracy, the
loops  are symmetric under  inversion of  the magnetization  and field
axis.  Note first that above  $T\simeq 43$ K no hysteresis effects are
observed,  consistently  with  the  behavior  of  $\chi(T)$  in  Fig.\
\ref{fig1}.  Secondly,  the loops in the  range from $15$ K  to $45$ K
are typical of a glassy system  while, at $T<15$ K the loops display a
jump associated with the  field induced ferromagnetic component.  This
becomes more pronounced  as the temperature is lowered.   On the other
hand, the virgin  curves do not show any  signature of a ferromagnetic
component, in  agreement with the behaviour of  the ZFC susceptibility
in Fig. \ref{fig1}.
\vspace{-9mm}
The characterization  of the onset  of the ferromagnetic  component is
presented  in Fig.\  \ref{fig3}, where  we  show the  behavior of  the
estimated  coercive  field  $H_{coe}(T)$  and the  magnetization  jump
$\Delta  M(T)$ as  a  function of  temperature \footnote{The  coercive
field is estimated by linear interpolation between the two data points
closer (above  and below)  to $H=0$.  $\Delta  M$ is estimated  as the
difference between  $M_+$ and $M_-$, that are  the extrapolated values
(through  a  parabolic  approximation  to  the $M(H)$  curve)  of  the
magnetization at  $H_{coe}(T)$.}.  The coercive field  exhibits a peak
at  $T_c \simeq 13$  K which,  within the  errors, coincides  with the
temperature where $\Delta M$ vanishes  and is close to the position of
the minimum  in the FC susceptibility.  These  results corroborate the
existence  of a continuous  phase transition  which occurs  within the
glassy phase.

\begin{center}
\begin{figure}[thb]
\includegraphics[width=8cm,clip]{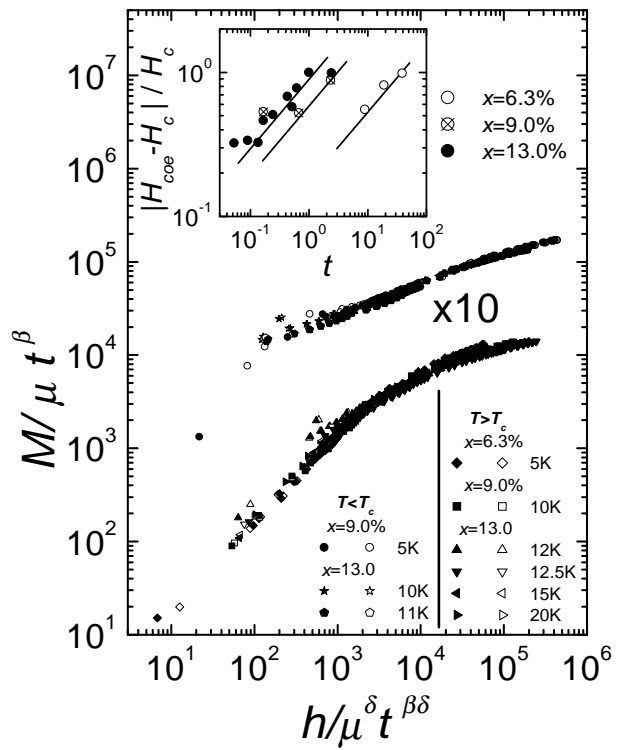}
\caption{\label{fig4} Scaling of the hysteresis loops corresponding to
different   temperatures  and   compositions  as   indicated   by  the
legend. The two sets of data corresponds to the $T>T_c(x)$ (lower) and
$T<T_c(x)$ (upper and multiplied by  a factor 10).  The inset shows in
log-log scale the behavior of $H_{coe}(x,T)$ when approaching $T_c(x)$
for different values  of $x$.  The lines have a  slope $\beta \delta =
0.4$. Solid  and open  symbols stand for  the ${\cal M}_+$  and ${\cal
M}_-$ branches respectively.}
\end{figure}
\end{center}

In order to understand the role played by the configurational disorder
and/or  thermal fluctuations  in this  transition it  is  important to
study  results  corresponding  to  other compositions  for  which  the
different stoichiometry  give rise  to different degrees  of disorder.
Besides the measurements  presented above corresponding to $x=13.0\%$,
we  have used  data  for polycrystalline  samples  with $x=6.3\%$  and
$x=9.0\%$,  published previously \cite{Obrado1999b}.   To characterize
the  critical behavior  we have  performed a  scaling analysis  of the
$M(H)$ curves.   The direct fitting  of the $\beta$ exponent  from the
$\Delta  M$ vs.   $T$  data  (in Fig.   \ref{fig3})  does not  provide
reliable  estimates of  $\beta$ since  the  values of  $\Delta M$  are
strongly   influenced  by   the  numerical   procedure  used   in  the
extrapolations  \footnotemark[13].   For   a  fixed  Mn  content,  the
distance  to  the  critical  point  is measured  through  
\vspace{-9mm}
the  reduced  temperature   $t=(T-T_c(x))/T_c(x)$  and  reduced  field
$h=(H-H_{coe}(x,T))/H_{coe}(x,T)$.   The  standard scaling  hypothesis
\cite{Sethna1993, Berger2000} for the positive and negative hysteresis
loop branches is $M_{\pm}= t^{\beta} {\cal M}_{\pm} \left ( h/t^{\beta
\delta} \right )$.  In order to scale data for different compositions,
we generalize  the preceding expression  to include the fact  that the
magnetic moment  $\mu(x)$ depends on the  Mn content.  Renormalization
of the magnetization by a factor $\mu(x)$ yields the following scaling
hypothesis:
\begin{equation}
M_{\pm}= \mu(x) t^{\beta}  {\cal M}_{\pm} \left ( h/ \mu(x)^{\delta}
t^{\beta \delta} \right ).
\label{scaling}
\end{equation}
This hypothesis is tested in Fig.\ref{fig4}, where we have represented
in a log-log scale $M/\mu t^\beta$ versus $h/\mu^\delta t^{\beta
\delta}$.  Data correspond to scaling of hysteresis loops at different
temperatures and compositions as indicated by the legend.  A very good
scaling of the curves is  obtained. The best data collapse corresponds
to $\beta=0.03\pm0.01$ and  $\beta \delta=0.4\pm0.1$.  The upper curve
in Fig.\ \ref{fig4}  (which has been shifted by a factor  of 10 in the
vertical  scale for clarity)  corresponds to  $T<T_c(x)$ data  and the
lower curve  to $T>T_c(x)$ data.   Solid (open) symbols  correspond to
${\cal M}_+$ (${\cal M}_-$) branches.  The values of $\mu(x)$ for each
of  the studied composition  are estimated  using the  derivative with
respect to $H$ of Eq.  (\ref{scaling}):
\begin{equation}
\mu(x) \propto \left [ \left ( \frac{\partial {\cal M}_{\pm}}{\partial
H}\right )_{H_{coe}} H_{coe} \right ]^{1/(1-\delta)} t^{-\beta}.
\end{equation}
The  obtained   values,  except  for  a   multiplicative  factor,  are
$\mu(x=6.3\%)=0.566$, $\mu(x=9.0\%)=0.575$, and $\mu(x=12.9\%)=0.625$.
The numerical  estimations of $T_c(x)$, obtained  from the improvement
of  the  collapses  in  Fig.  \ref{fig4},  are  $T_c(x=6.3\%)=0.5$  K,
$T_c(x=9.0\%)=6.0$ K  and $T_c(x=13.0\%)=13.2$  K. It is  important to
note  that  the collapse  of  the  data  corresponding to  $x=13.0\%$,
considered separately from the other compositions (without the need of
fitting $\mu$), render values of $\beta$ and $\beta \delta$ within the
same error bars.
\vspace{-9mm}
An independent  method to check the  consistency of the  scaling is to
analyze   the  critical  behavior   of  the   $H_{coe}(x,T)$  surface.
According  to the Clausius-Clapeyron  equation, assuming  the standard
scaling  behavior for  the  entropy and  the  magnetization and  using
Griffiths equality \cite{Stanley1971}, one obtains
\begin{equation}
\left  ( \frac{\partial  H_{coe}(x,T)}{ \partial  T}\right  )_{x} \sim
t^{\beta \delta -1}.
\label{clausius}
\end{equation}
Integration   yields   $[H_{coe}(x,T)-H_c(x)]/H_c(x)   \sim   t^{\beta
\delta}$.  The  consistency of the data with  this predicted power-law
behavior is  checked in  the inset of  Fig. \ref{fig4}.   The straight
lines have a slope of $\beta \delta = 0.4$. For the $x=13\%$ case, for
which the  amount of data  is larger, we  can perform a linear  fit
 of  such a behavior, rendering $H_c(x=13.0\%)  = 560 \pm
15$ Oe.  The fit is  shown in Fig.  \ref{fig3}. 

\begin{center}
\begin{figure}[thb]
\includegraphics[width=6cm,clip]{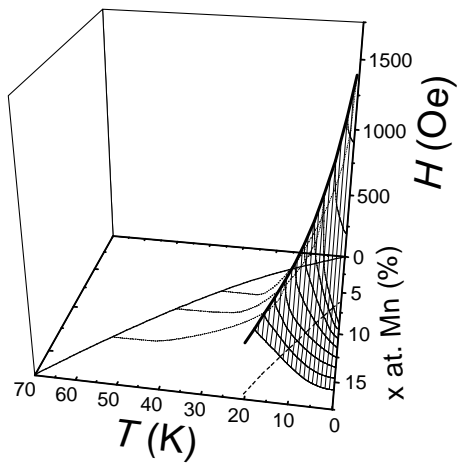}
\caption{\label{fig5} Full $(H,T,x)$ phase diagram. The thick continuous
line  indicates the  second  order disorder  induced phase  transition
whose projection in  the $(x,T)$ plane is shown by  the dashed line. The
continuous  line  across the  $(x,T)$  plane  is  an estimation  of  the
freezing  transition, where  metastable hysteresis  phenomena appears.
The thin dotted lines indicate the behavior of $H_{coe}$ for different
values  of $x$  for  $T>T_c(x)$. The  grid  indicates the  first-order
transition surface where discontinuity of the magnetization occurs.}
\end{figure}
\end{center}

Fig.\  \ref{fig5}  shows   the  complete  $(H,T,x)$  metastable  phase
diagram.  The  line across the $(x,T)$  plane is an  estimation of the
freezing  temperature obtained  from the  susceptibility  maxima (from
Fig.\ref{fig1}  and from  Ref.  \onlinecite{Obrado1999b}).  To  a good
approximation,   this  temperature   determines   the  spin   freezing
temperature   in   the  low   concentration   region.   However,   for
concentrations larger than $8 \%$, as 
\vspace{-9mm} can  be seen  in Fig.\ \ref{fig1},  the position  of the
peak overestimates the  temperature where irreversibility occurs.  The
thick  line  is the  line  of  critical points,  at  the  edge of  the
first-order  transition  surface   (shaded)  where  the  magnetization
exhibits  a discontinuity.  The  projection of  this critical  line is
also shown on the $H=0$ plane.

The exponents $\beta$ and $\beta  \delta$ obtained in the present work
are in  good agreement  with those reported  recently in the  study of
Co/CoO bilayers  \cite{Berger2000}.  In  that case, the  authors argue
that their  system exhibits a disorder induced  phase transition.  The
amount of  disorder is controlled through  the antiferromagnetic order
of the CoO layer which changes when the temperature crosses the N\'eel
point. In  our case we  can guess a  similar origin, since it  is know
that     ferromagnetism    and    antiferromagnetism     coexist    in
non-stoichiometric                    Heusler                   alloys
\cite{Prado1998,Obrado1999,Acet2002}.    This   hypothesis   is   also
consistent  with the  behavior  of the  hysteresis  loops that  become
closed  at relatively  small fields  but do  not saturate  up  to high
fields.  We suggest  that even at the low  temperatures studied in the
present  work, the  antiferromagnetically interacting  moments  act as
random  fields  on  the  frozen  ferromagnetic  moments.   Temperature
changes  are  able  to  alter  this  effective  disorder  causing  the
disorder-induced  phase transition.   We want  to note  that  the zero
temperature model for Cu-Al-Mn \cite{Obrado1999b}, which considers the
random position of Mn atoms on the four interpenetrated sublattices in
the fcc structures and the  coupling through an RKKY interaction, also
predicts  that such  a system  will display  a disorder  induced phase
transition at $T=0$ and $x=7.5\%$. This is very close to the estimated
value of  $x$ for which  $T_c\rightarrow 0$.  As occurs  with previous
experimental estimations for Co/CoO bilayers, the low value of $\beta$
is  in agreement with  the predicted  theoretical exponents  for short
ranged zero-temperature  3d models with  metastable dynamics. However,
the  value of  $\beta  \delta$ falls  out  of the  error  bars of  the
numerical  estimations.   It is  worth  to  note  that Berger  et  al.
claimed  that their  system, in  contrast to  our's, behaved  as  a 2d
system. This is, however, in  contradiction with the fact that for the
2d  case,  no  disorder  induced  phase  transition  is  theoretically
expected \cite{Perkovic1996}.

In  conclusion we  have proven  the  existence of  criticality in  the
magnetic hysteresis  loops in Cu-Al-Mn.   The non-equilibrium critical
line has  been determined in the $(H,T,x)$  diagram.  Both temperature
and  composition are  the variables  that allow  to tune  the critical
amount of  disorder.  The  coexistence of ferro  and antiferromagnetic
components in the system is at the origin of the observed behaviour.

This  work  has received  financial  support  from  CICyT (Project  No
MAT2001-3251),  CIRIT  (Project  2001SGR00066)  and  IBERDROLA.   J.M.
acknowledges financial support from DURSI.


\begin{thebibliography}{12}
\expandafter\ifx\csname natexlab\endcsname\relax\def\natexlab#1{#1}\fi
\expandafter\ifx\csname bibnamefont\endcsname\relax
  \def\bibnamefont#1{#1}\fi
\expandafter\ifx\csname bibfnamefont\endcsname\relax
  \def\bibfnamefont#1{#1}\fi
\expandafter\ifx\csname citenamefont\endcsname\relax
  \def\citenamefont#1{#1}\fi
\expandafter\ifx\csname url\endcsname\relax
  \def\url#1{\texttt{#1}}\fi
\expandafter\ifx\csname urlprefix\endcsname\relax\def\urlprefix{URL }\fi
\providecommand{\bibinfo}[2]{#2}
\providecommand{\eprint}[2][]{\url{#2}}

\bibitem[{\citenamefont{Sethna et~al.}(1993)\citenamefont{Sethna, Dahmen,
  Kartha, Krumhansl, Roberts, and Shore}}]{Sethna1993}
\bibinfo{author}{\bibfnamefont{J.~P.} \bibnamefont{Sethna}},
  \bibinfo{author}{\bibfnamefont{K.}~\bibnamefont{Dahmen}},
  \bibinfo{author}{\bibfnamefont{S.}~\bibnamefont{Kartha}},
  \bibinfo{author}{\bibfnamefont{J.~A.} \bibnamefont{Krumhansl}},
  \bibinfo{author}{\bibfnamefont{B.~W.} \bibnamefont{Roberts}},
  \bibnamefont{and} \bibinfo{author}{\bibfnamefont{J.~D.} \bibnamefont{Shore}},
  \bibinfo{journal}{Phys.\ Rev.\ Lett.} \textbf{\bibinfo{volume}{70}},
  \bibinfo{pages}{3347} (\bibinfo{year}{1993}).

\bibitem[{\citenamefont{Vives and Planes}(1994)}]{Vives1994}
\bibinfo{author}{\bibfnamefont{E.}~\bibnamefont{Vives}} \bibnamefont{and}
  \bibinfo{author}{\bibfnamefont{A.}~\bibnamefont{Planes}},
  \bibinfo{journal}{Phys.\ Rev. B} \textbf{\bibinfo{volume}{50}},
  \bibinfo{pages}{3839} (\bibinfo{year}{1994}).

\bibitem[{\citenamefont{E.Vives and A.Planes}(2001)}]{Vives2001}
\bibinfo{author}{\bibnamefont{E.Vives}} \bibnamefont{and}
  \bibinfo{author}{\bibnamefont{A.Planes}}, \bibinfo{journal}{Phys. Rev. B}
  \textbf{\bibinfo{volume}{63}}, \bibinfo{pages}{134431}
  (\bibinfo{year}{2001}).

\bibitem[{\citenamefont{Perkovi{\'c} et~al.}(1999)\citenamefont{Perkovi{\'c},
  Dahmen, and J.P.Sethna}}]{Perkovic1999}
\bibinfo{author}{\bibfnamefont{O.}~\bibnamefont{Perkovi{\'c}}},
  \bibinfo{author}{\bibfnamefont{K.~A.} \bibnamefont{Dahmen}},
  \bibnamefont{and} \bibinfo{author}{\bibnamefont{J.P.Sethna}},
  \bibinfo{journal}{Phys.\ Rev.\ B} \textbf{\bibinfo{volume}{59}},
  \bibinfo{pages}{6106} (\bibinfo{year}{1999}).

\bibitem[{\citenamefont{Dahmen and J.P.Sethna}(1996)}]{Dahmen1996}
\bibinfo{author}{\bibfnamefont{K.~A.} \bibnamefont{Dahmen}} \bibnamefont{and}
  \bibinfo{author}{\bibnamefont{J.P.Sethna}}, \bibinfo{journal}{Phys.\ Rev. B}
  \textbf{\bibinfo{volume}{53}}, \bibinfo{pages}{14872} (\bibinfo{year}{1996}).

\bibitem[{\citenamefont{E.Obrad\'o
  et~al.}(1999{\natexlab{a}})\citenamefont{E.Obrad\'o, E.Vives, and
  A.Planes}}]{Obrado1999}
\bibinfo{author}{\bibnamefont{E.Obrad\'o}},
  \bibinfo{author}{\bibnamefont{E.Vives}}, \bibnamefont{and}
  \bibinfo{author}{\bibnamefont{A.Planes}}, \bibinfo{journal}{Phys.\ Rev.\ B}
  \textbf{\bibinfo{volume}{59}}, \bibinfo{pages}{13901}
  (\bibinfo{year}{1999}{\natexlab{a}}).

\bibitem[{\citenamefont{A.Berger et~al.}(2000)\citenamefont{A.Berger,
  A.Inomata, J.S.Jiang, J.E.Pearson, and S.D.Bader}}]{Berger2000}
\bibinfo{author}{\bibnamefont{A.Berger}},
  \bibinfo{author}{\bibnamefont{A.Inomata}},
  \bibinfo{author}{\bibnamefont{J.S.Jiang}},
  \bibinfo{author}{\bibnamefont{J.E.Pearson}}, \bibnamefont{and}
  \bibinfo{author}{\bibnamefont{S.D.Bader}}, \bibinfo{journal}{Phys. Rev.
  Lett.} \textbf{\bibinfo{volume}{85}}, \bibinfo{pages}{4176}
  (\bibinfo{year}{2000}).

\bibitem[{\citenamefont{E.Obrad\'o
  et~al.}(1999{\natexlab{b}})\citenamefont{E.Obrad\'o, A.Planes, and
  B.Mart\'{\i}nez}}]{Obrado1999b}
\bibinfo{author}{\bibnamefont{E.Obrad\'o}},
  \bibinfo{author}{\bibnamefont{A.Planes}}, \bibnamefont{and}
  \bibinfo{author}{\bibnamefont{B.Mart\'{\i}nez}}, \bibinfo{journal}{Phys.\
  Rev.\ B} \textbf{\bibinfo{volume}{59}}, \bibinfo{pages}{11450}
  (\bibinfo{year}{1999}{\natexlab{b}}).

\bibitem[{\citenamefont{H.E.Stanley}(1983)}]{Stanley1971}
\bibinfo{author}{\bibnamefont{H.E.Stanley}}, \emph{\bibinfo{title}{Introduction
  to Phase Transitions and Critical Phenomena}}, vol.~\bibinfo{volume}{8} of
  \emph{\bibinfo{series}{International Series of Monographs on Physics}}
  (\bibinfo{publisher}{Oxford University Press}, \bibinfo{address}{New York},
  \bibinfo{year}{1983}).

\bibitem[{\citenamefont{M.O.Prado et~al.}(1998)\citenamefont{M.O.Prado,
  F.C.Lovey, and L.Civale}}]{Prado1998}
\bibinfo{author}{\bibnamefont{M.O.Prado}},
  \bibinfo{author}{\bibnamefont{F.C.Lovey}}, \bibnamefont{and}
  \bibinfo{author}{\bibnamefont{L.Civale}}, \bibinfo{journal}{Acta Mater.}
  \textbf{\bibinfo{volume}{46}}, \bibinfo{pages}{137} (\bibinfo{year}{1998}).

\bibitem[{\citenamefont{M.Acet et~al.}(2002)\citenamefont{M.Acet, E.Duman,
  E.Wasserman, Ll.Ma{\~n}osa, and A.Planes}}]{Acet2002}
\bibinfo{author}{\bibnamefont{M.Acet}},
  \bibinfo{author}{\bibnamefont{E.Duman}},
  \bibinfo{author}{\bibnamefont{E.Wasserman}},
  \bibinfo{author}{\bibnamefont{Ll.Ma{\~n}osa}}, \bibnamefont{and}
  \bibinfo{author}{\bibnamefont{A.Planes}}, \bibinfo{journal}{J. Appl. Phys.}
  \textbf{\bibinfo{volume}{92}},  (\bibinfo{year}{2002}).

\bibitem[{\citenamefont{O.Perkovi{\'c}
  et~al.}(1996)\citenamefont{O.Perkovi{\'c}, K.A.Dahmen, and
  J.P.Sethna}}]{Perkovic1996}
\bibinfo{author}{\bibnamefont{O.Perkovi{\'c}}},
  \bibinfo{author}{\bibnamefont{K.A.Dahmen}}, \bibnamefont{and}
  \bibinfo{author}{\bibnamefont{J.P.Sethna}},
  \bibinfo{journal}{cond-mat/9609072}  (\bibinfo{year}{1996}).

\end{thebibliography}
\end{document}